\documentclass[12pt]{article}
\usepackage{amsmath}
\usepackage{color}
\usepackage{graphicx}
\begin{document}
\baselineskip=18 pt
\begin{center}
{\large{\bf Axial symmetry Type N space-time with a naked curvature singularity and Closed Time-like Curves}}
\end{center}

\vspace{.5cm}

\begin{center}
{\bf Faizuddin Ahmed}\footnote{faizuddinahmed15@gmail.com ; faiz4U.enter@rediffmail.com}\\
{\bf Ajmal College of Arts and Science, Dhubri-783324, Assam, India}
\end{center}

\vspace{.5cm}

\begin{abstract}

A family of type N exact solution of the Einstein's field equations, regular everywhere except on the symmetry axis where it possesses a naked curvature singularity, is present. The stress-energy tensor is of the anisotropic fluid coupled with pure radiation field satisfy the different energy conditions and the physical parameters diverge at $r \rightarrow 0$. The space-time admits a non-expanding, non-twisting, and shear-free geodesic null congruence and belongs to a special class of type N Kundt metrics. The space-time is geodesically complete along the radial direction in the constant $z$-planes and exhibits geometrically different properties from the known {\it pp}-waves. The present family of solution admits closed time-like curves (CTC) which appear after a certain instant of time and the space-time is a four-dimensional generalization of the Misner space metric in curved space-time.

\end{abstract}

{\bf Keywords:} exact solutions, stress-energy tensor, naked singularity, Kundt metrics, closed time-like curves, energy conditions.
\vspace{0.3cm}

{\bf PACS numbers:} 04.20.Jb, 04.20.Gz, 04.20.-q, 04.20.Dw

\vspace{.5cm}

\section{Introduction}

All vacuum or non-vacuum solutions of the Einstein's field equations with or without cosmological constant, which are of type N with non-twisting, and shear-free geodesic null congruence have been extensively studied (see, Refs. \cite{PTEP3,EPJC} and references therein). A complete class of non-twisting type N vacuum solution with non-zero cosmological constant was found in Ref. \cite{Garcia}. These solutions are further analyzed in Ref. \cite{Ozvath} and subsequently by various authors (see, Refs.\cite{Bicak,Bicak2,Podo,Bergh,Hazara,Ort}). Type N Kundt metrics \cite{Kundt1,Kundt2} are the general class of vacuum or non-vacuum solution to the Einstein's field equations which admit a non-expanding, non-twisting, and shear-free geodesic null congruence. If this shear-free geodesics null congruence is a covariantly constant vector field, we called this solution as {\it pp}-wave or plane waves space-time ({\it e. g.}, Refs. \cite{GERG,AOP,PTEP,TMPH,AOP3} and references therein). In Ref. \cite{Siklos}, a special class of type N Kundt metrics of the non-zero cosmological constant, non-vacuum solution to the field equations which exhibit geometrically different properties from the known {\it pp}-waves was obtained. Subsequently, a type N vacuum solution of non-zero cosmological constant, special sub-case of the Siklos class of solution, was obtained in Ref. \cite{Kai}. Recently, a family of type N aligned pure radiation field solution of non-zero cosmological constant was constructed in Ref. \cite{PTEP3}. This family of solution admits closed time-like curves and exhibit geometrically different properties from the known {\it pp}-waves space-time. Note that this family of solution  \cite{PTEP3} is different from the Siklos class of solution \cite{Siklos} in the sense that the first family of solution admits closed time-like curves which  violate the causality condition. The type N vacuum solution of non-zero cosmological constant obtained in Ref. \cite{AOP2}, and pure radiation field type N solution of the non-zero cosmological constant in Ref. \cite{EPJC} are the special sub-case of this family of solution \cite{PTEP3}. It is worth mentioning that all the above type N vacuum or non-vacuum solutions are free from curvature singularities. Few of them violate the causality condition and some others are not.

However, some other algebraically special solution showing geometrically different properties from {\it pp}-waves or plane-waves possess a naked curvature singularity and of these, few solutions violate the causality condition. This type of solution with a naked curvature singularity includes a type D vacuum space-time in Ref. \cite{AHEP}, axial symmetric type II null dust fluid space-time in Ref. \cite{AHEP2}, cylindrical symmetric type D non-vacuum space-time \cite{AHEP3}, type N non-vacuum space-time in Ref. \cite{PTEP2}, and axial symmetry type II space-time in Ref. \cite{IJGMMP} which are violate the causality condition. Other investigation of gravitational collapse solution with a naked singularity satisfying the causality condition includes counter-rotating dust shell cylinder \cite{Apo}, the evolution of cylindrical dust shell \cite{Eche}, the collapse of non-rotating and infinite dust cylinder \cite{Gutt}, the high-speed collapse of cylindrical symmetry thick shell model composed of dust \cite{Nakao}, perfect fluid solution with non-vanishing pressure \cite{Nakao1}, cylindrical symmetric collapse of counter-rotating dust shells \cite{Gonc,Nolan,Peri}, self-similar scalar field \cite{Wang,Wang2}, cylindrical symmetry conformally flat anisotropic space-time \cite{AHEP4}, cylindrical symmetric type D vacuum space-time \cite{EPJA}, four-dimensional non-static space-time \cite{EPJA2}, and cylindrical symmetric, non-rotating and non-static or static black hole solutions and the naked singularities \cite{EPJC2}. Examples of non-spherical gravitational collapse solution with a naked singularity would be \cite{Chi,Na,Piran,JMM,Bondi,Th,Morgan,Lete,M,M1}. To counter the occurrence of naked singularity in a solution of Einstein's field equations, R. Penrose proposed the Cosmic Censorship Conjecture (CCC) \cite{Pen1,Pen2,Pen3}. However, there is no theorem or proof yet has been known which support or counter this Conjecture. On the contrary, there are no mathematical details yet known which forbid the appearance of a naked singularity in a solution of the field equations.

In this paper, we attempt to construct a family of zero cosmological constant type N solution to the field equations with a naked curvature singularity. This family of solution exhibit geometrically different properties from the known {\it pp}-waves and violate the causality condition. We show the family of type N solution study here is a four-dimensional generalization of the Misner space metric in curved space-time. 

\section{A four-dimensional curved space-time}

We consider the following time-dependent four-dimensional rotating curved space-time given by 
\begin{equation}\label{eq1}
ds^2=g_{rr}\,dr^2+g_{zz}\,dz^2+2\,g_{t\psi}\,dt\,d\psi+g_{\psi\psi}\,d\psi^2+2\,g_{r\psi}\,dr\,d\psi+2\,g_{z\psi}\,dz\,d\psi,
\end{equation}
where the metric functions are as follow :
\begin{equation}\label{eq2}
\begin{split}
g_{rr}&= g_{rr} (r),\\
g_{zz}&= g_{zz} (r),\\
g_{\psi\psi}&= g_{\psi\psi} (t,r),\\
g_{t\psi}&= g_{t\psi} (t,r),\\
g_{r\psi}&= g_{r\psi} (r),\\
g_{z\psi}&= g_{z\psi} (r,z)=z\,g_{z\psi} (r).
\end{split}
\end{equation}
The ranges of the coordinates are 
\begin{equation}\label{eq3}
-\infty < t < \infty,\quad 0 \leq r <\infty,\quad -\infty < z < \infty
\end{equation}
and the coordinate $\psi$ is chosen cyclic (and/or periodic locally), that is each $\psi$ is identified with $\psi+\psi_0$ for a certain parameter $\psi_0>0$ or $0 \leq \psi \leq \psi_0$ \cite{Ori} (see also, Refs. \cite{AHEP,AHEP2}). For the space-time study here, we have assumed that $\boldsymbol{X}\rightarrow 0$ as $r \rightarrow 0$, where  $\boldsymbol{X}=|\eta^{\mu}\,\eta^{\nu}\,g_{\mu\nu}|=|g_{\psi\psi}|$ and $\partial_{\psi}$ is a spacelike killing vector \cite{Steph}. However, for the chosen  metric functions (\ref{eq2}) the space-time (\ref{eq1}) fails to satisfy the elementary flatness condition \cite{Steph} which states that in the limit of rotation axis $r\rightarrow 0$, \begin{equation}
    \frac{g^{\mu\nu}\,(\nabla_{\mu} \boldsymbol{X})\,(\nabla_{\nu} \boldsymbol{X})}{4\, \boldsymbol{X}} \rightarrow 1.
    \label{elementary}
\end{equation}
Therefore, the symmetry axis $r=0$ is singular or non-regular which is not well-defined. The singular axis may contain some line-like sources and/or exist conical singularities \cite{Steph}. The appearance of space-time singularities on the axis would be considered as representing the existence of some kind of sources \cite{BB,BB2,Sil,SH}.

From the above family of solution (\ref{eq1}), one can construct a family of type N solution with or without a cosmological constant. A family of type N aligned pure radiation field solution with non-zero cosmological constant admitting closed time-like curves was constructed in \cite{PTEP3}. In this work, we attempt to construct a family of type N solution especially non-vacuum solution of the field equations with zero cosmological constant violating the causality condition possesses a naked singularity.

We choose the following metric functions for zero cosmological constant non-vacuum solution in the metric (\ref{eq1}):
\begin{equation}\label{eq4}
\begin{split}
g_{rr} (r)&= A (r),\\
A (r)&= [B'(r)]^2\,B (r),\\
g_{\psi\psi} (t,r)&= -a(t)\,B (r),\\
g_{t\psi} (t,r)&= -c_{0}\,{\dot a}(t)\,B (r),\\
g_{r\psi} (r)&= c_{2}\,C (r),\\
g_{z\psi} (r)&= c_{1}\,B (r),\\
g_{zz} (r)&= B (r).
\end{split}
\end{equation}
Here prime denotes ordinary derivative w. r. t. $r$, $c_0>0, c_1>0, c_2>0$ are constants, $B(r), C(r)$ are free functions and dot stands for ordinary derivative w. r. t. $t$. The space-time (\ref{eq1}) with (\ref{eq4}) can be express as
\begin{eqnarray}\label{eq5}
ds^2&=&B (r)\,[B'^{2} (r)\,dr^2+dz^2-2\,c_{0}\,\dot{a} (t)\,dt\,d\psi-a (t)\,d\psi^2+2\,c_{1}\,z\,dz\,d\psi]\nonumber\\
&&+2\,c_{2}\,C (r)\,dr\,d\psi.
\end{eqnarray}
Replacing the following
\begin{equation}\label{eq6}
r\rightarrow {\bar r}:=B (r)\quad,\quad t\rightarrow T:= a (t)
\end{equation}
into the above metric (\ref{eq5}) and a redefinition of $C (r)$ by $H (r)$, we get a family of $\Lambda=0$ type N solution (finally dropping bar)
\begin{equation}
ds^2=r\,[dr^2+dz^2-2\,c_{0}\,dT\,d\psi-T\,d\psi^2+2\,c_{1}\,z\,dz\,d\psi+2\,c_{2}\,H (r)\,dr\,d\psi].
\label{eq7}
\end{equation}
The solution (\ref{eq7}) is our main aim that we shall discuss below in details.

The above metric has signature $(-,+,+,+)$ and the coordinates are label as $x^0=T$, $x^1=r$, $x^2=\psi$, $x^3=z$ and the determinant of the metric tensor $g_{\mu\nu}$ is
\begin{equation}\label{eq8}
det\;g=-c^{2}_{0}\,r^4.
\end{equation}
The non-zero components of the Einstein tensor $G^{\mu}_{\nu}$ ($\mu,\nu=0,1,2,3)$ are
\begin{equation}\label{eq9}
\begin{split}
G^{0}_{0}&= -G^{1}_{1}=G^{2}_{2}=G^{3}_{3}=-\frac{3}{4\,r^3},\nonumber\\
G^{0}_{1}&= \frac{3\,c_2\,H(r)}{2\,c_0\,r^3},\\
G^{0}_{2}&= -\frac{1}{2\,c^{2}_{0}\,r^2}\,[c_1\,r+c_2\,\{ H(r)+r\,H' (r) \}].
\end{split}
\end{equation}
We have calculated the scalar curvature invariant for the metric (\ref{eq7}). These are given by
\begin{equation}\label{eq100}
\begin{split}
R^{\mu}_{\,\mu}&= \frac{3}{2\,r^3},\\
R^{\mu\nu}\,R_{\mu\nu}&= \frac{9}{4\,r^6},\\
R^{\mu\nu\rho\sigma}\,R_{\mu\nu\rho\sigma}&= \frac{15}{4\,r^3}
\end{split}
\end{equation}
which diverges at $r=0$ not covered by an event horizon. In addition, these scalar invariant vanish rapidly at spatial infinity ( $r\rightarrow \infty$) along the radial direction. Thus the solution (\ref{eq7}) study here is asymptotically flat along the radial direction and possesses a naked curvature singularity and hence the cosmic censorship conjecture doesn't hold good here.

The only non-zero Weyl tensor $C_{\mu\nu\rho\sigma}$ of the  solution study here is
\begin{equation}
C_{1212}=C_{r\psi r\psi}=\frac{r}{4\,c_0}\,(-c_1+c_2\,H'(r)).
\label{300}
\end{equation}
Note that if one sets $H(r)=\frac{c_1}{c_2}\,r$, then the  solution (\ref{eq7}) represents a conformally flat type O metric with a naked singularity.

\subsection{\bf Stress-energy tensor and the Energy conditions}

The Einstein's field equations with $\Lambda=0$ are given by
\begin{equation}\label{eq10}
G^{\mu\nu}=R_{\mu\nu}-\frac{1}{2}\,R\,g_{\mu\nu}=T^{\mu\nu},\quad \mu,\nu=0,1,2,3,
\end{equation}
where $T^{\mu\nu}$ is the stress-energy tensor, $R_{\mu\nu}$ is the Ricci tensor, $R$ is the Ricci scalar and units are taken $8\,\pi\,G=1=c=\hbar$. 

We choose the stress-energy tensor is of pure radiation field coupled with the anisotropic fluid given by
\begin{equation}\label{eq11}
T^{\mu\nu}=\mu\,k^{\mu}\,k^{\nu}+(\rho+p_{t})\,U^{\mu}\,U^{\nu}+p_{t}\,g^{\mu\nu}+(p_{r}-p_{t})\,\zeta^{\mu}\,\zeta^{\nu},
\end{equation}
where $U^{\mu}$ is the unit time-like four-velocity vector, $\zeta^{\mu}$ is the unit spacelike vector along the radial direction $r$, and $k^{\mu}$ is a null vector field which satisfies the following relation :
\begin{equation}\label{eq12}
U^{\mu}\,k_{\mu}=0=U^{\mu}\,\zeta_{\mu},\quad U^{\mu}\,U_{\mu}=-1,\quad k^{\mu}\,\zeta_{\mu}=0,\quad \zeta_{\mu}\,\zeta^{\mu}=1,\quad k^{\mu}\,k_{\mu}=0.
\end{equation}
Here $\mu$ is the radiation energy density, and $\rho$, $p_r$, $p_t$ are the fluid energy density, the radial pressure and tangential pressure, respectively.

The null vector field $k^{\mu}$ is the tangent vector field of geodesics null congruence the radiation propagates along and satisfies the following condition :
\begin{equation}\label{eq13}
k_{\mu\,;\,\nu}\,k^{\nu}=0.
\end{equation}
The Ricci scalar from the field equations (\ref{eq10}) using (\ref{eq11})-(\ref{eq12}) is
\begin{equation}\label{eq14}
-R=T=2\,p_{t}+p_{r}-\rho.
\end{equation}
For the metric (\ref{eq7}), we define the time-like unit four veclocity vector $U^{\mu}$
\begin{equation}\label{eq15}
U^{\mu}=\frac{1}{\sqrt{2}}\,(f(r,t), 0, -1, 0),\quad U^{\mu}\,U_{\mu}=-1,
\end{equation}
where
\begin{equation}\label{eq16}
f(r,t)=-\frac{1}{c_0\,r}+\frac{t}{2\,c_0}.
\end{equation}
The spacelike unit vector $\zeta_{\mu}$ along $r$ and the null vector $k_{\mu}$ are define by
\begin{equation}\label{eq17}
\zeta_{\mu}=(0,\sqrt{r},0,0)=\sqrt{r}\,\delta^{r}_{\mu},\quad k_{\mu}=(0,0,1,0)=\delta^{\psi}_{\mu}.
\end{equation}

From the stress-energy tensor (\ref{eq11}) using (\ref{eq15})---(\ref{eq17}), we get 
\begin{equation}\label{eq18}
\begin{split}
T^{0}_{0}&= \frac{1}{4}\,\{2\,(p_{t}-\rho)+(p_{t}+\rho)\,r\,t\},\\
T^{0}_{1}&= -\frac{c_2\,H (r)}{4\,c_0}\,\{-4\,p_{r}+2\,(p_{t}-\rho)+(p_{t}+\rho)\,r\,t \},\\
T^{0}_{2}&= \frac{1}{8\,c_0\,r}\,\{-8\,\mu+(-4+r^2\,t^2)\,(p_{t}+\rho)\},\\
T^{0}_{3}&= -\frac{c_1}{4\,c_0}\,z\,(-2+r\,t)\,(p_{t}+\rho),\\
T^{1}_{1}&= p_{r},\\
T^{2}_{0}&= -\frac{1}{2}\,c_0\,r\,(p_{t}+\rho),\\
T^{2}_{1}&= \frac{1}{2}\,c_2\,r\,(p_{t}+\rho)\,H (r),\\
T^{2}_{2}&= \frac{1}{4}\,\{2\,(p_{t}-\rho)-(p_{t}+\rho)\,r\,t\},\\
T^{2}_{3}&= \frac{1}{2}\,c_1\,r\,z\,(p_{t}+\rho),\\
T^{3}_{3}&= p_{t}.
\end{split}
\end{equation}

From the field equations (\ref{eq10}) using Eqs. (\ref{eq9}), (\ref{eq18}) and after simplification, we have the following non-zero physical parameters
\begin{equation}\label{eq19}
\rho=p_{r}=-p_{t}=\frac{3}{4\,r^3},\quad \mu=\frac{1}{2\,c_0}\,[c_1+c_2\,\{ \frac{H (r)}{r}+H' (r) \}].
\end{equation}
The stress-energy tensor satisfies the following energy condition \cite{Hawking}
\begin{equation}\label{eq20}
\begin{split}
\boldsymbol{WEC}\,&:  \quad \rho>0,\quad \mu>0,\\
\boldsymbol{WEC}_{t}\,&: \quad  \rho+p_{t}=0,\\
\boldsymbol{WEC}_{r}\,&: \quad  \rho+p_{r}>0,\\
\boldsymbol{SEC}\,&: \quad \rho+p_{r}+2\,p_{t}=0,\\
\boldsymbol{DEC}\,&: \quad \rho=|p_{i}|,\quad i=t,r,z.
\end{split}
\end{equation}
The physical parameters ($\rho$, $p_{r}$, $p_{t}$) are singular on the symmetry axis $r=0$. Therefore, the stress-energy tensor the anisotropic fluid diverge on the symmetry axis, and thus a naked singularity is formed.

Few examples of this family of type N solution are as follow:

\vspace{0.1cm}
{\bf Example 1}: If one choose the function $H(r)=\frac{c_3}{r}$ in the metric (\ref{eq7}), where $c_3>0$.
\vspace{0.1cm}

In that case the solution (\ref{eq7}) represents a Petrov type N solution of anisotropic fluid coupled with constant energy-density radiation field. The different physical parameters are
\begin{equation}\label{eq21}
\rho=p_{r}=-p_{t}=\frac{3}{4\,r^3},\quad \mu=\frac{c_1}{2\,c_0}.
\end{equation}
Recently, the author constructed a type N solution of anisotropic fluid coupled with radiation field \cite{PTEP2}, a special sub-case of the present family of type N solution (\ref{eq7}).

\vspace{0.1cm}
{\bf Example 2}: If one choose the function $H(r)=-\frac{c_1\,r}{2\,c_2}$ in the metric (\ref{eq7}).
\vspace{0.1cm}

In that case the solution (\ref{eq7}) represents Petrov type N solution of only anisotropic fluid. The physical parameters
\begin{equation}\label{eq22}
\rho=p_{r}=-p_{t}=\frac{3}{4\,r^3}
\end{equation}
diverge on the symmetry axis $r=0$.

\vspace{0.1cm}
{\bf Example 3}: If one choose the function $H(r)=\frac{c_3}{r}-\frac{c_1\,r}{2\,c_2}$ in the metric (\ref{eq7}), where $c_3>0$.
\vspace{0.1cm}

In that case also the solution (\ref{eq7}) represents Petrov type N solution of only anisotropic fluid with the physical parameters (\ref{eq22}).

\vspace{0.1cm}
{\bf Example 4}: If one choose the function $H(r)=\frac{c_1\,r}{c_2}$ in the metric (\ref{eq7}).
\vspace{0.1cm}

In that case the solution (\ref{eq7}) represents a conformally flat type $O$ solution of anisotropic fluid coupled with constant energy-density radiation field. The physical parameters of fluids are
\begin{equation}\label{eq23}
\rho=p_{r}=-p_{t}=\frac{3}{4\,r^3}\quad,\quad \mu=\frac{3\,c_1}{2\,c_0}.
\end{equation}

\subsection{\bf Geodesic analysis and strength of the naked singularities:}

In this sub-section, we focus on the radial geodesics which necessarily hit the singularity $r=0$ \cite{Kurtia} and finally strength of the naked singularities.

The Lagrangian for the metric (\ref{eq7}) is given by
\begin{equation}\label{eq24}
\begin{split}
L&= \frac{1}{2}\,g_{\mu\nu}\,\dot{x}^{\mu}\,\dot{x}^{\nu}\\
&=  \frac{1}{2}\,r\,[\dot{r}^2+\dot{z}^2-T\,\dot{\psi}^2-2\,c_0\,\dot{T}\,\dot{\psi}+2\,c_1\,z\,\dot{z}\,\dot{\psi}+2\,c_2\,H(r)\,\dot{r}\,\dot{\psi}],
\end{split}
\end{equation}
where dot stands derivative w. r. t. an affine parameter $\lambda$. From (\ref{eq15}), it is clear that $\psi$ is a cyclic coordinate. There exist constant of motion corresponding to this cyclic coordinate, {\it i. e.,} the azimuthal angular momentum $p_{\psi}$ which is a constant given by
\begin{equation}\label{eq25}
p_{\psi}=const=-T\,\dot{\psi}-c_0\,\dot{T}+c_1\,z\,\dot{z}+c_2\,H(r)\,\dot{r}.
\end{equation}

For the metric (\ref{eq7}), the geodesic equation for $T$, $r$ coordinates in explicit form are 
\begin{equation}\label{eq26}
\begin{split}
\ddot{T}&= \frac{1}{2\,c^{2}_0\,r}\,[-r\,T\,\dot{\psi}^2-c^{2}_{1}\,r\,z^2\,\dot{\psi}^2+c_2\,H^{2} (r)\,\dot{\psi}\,(2\,c_0\,\dot{r}-r\,\dot{\psi})+2\,c_0\,c_1\,r\,\dot{z}^2\\
&- 2\,c^{2}_{0}\,\dot{r}\,\dot{T}-2\,c_{0}\,r\,\dot{\psi}\,\dot{T}+c_0\,c_2\,H (r)\,\{\dot{r}^2-T\,\dot{\psi}^2+2\,c_1\,z\,\dot{\psi}\,\dot{z}+\dot{z}^2-2\,c_0\,\dot{\psi}\,\dot{T}\}\\
&+ 2\,c_0\,c_2\,r\,\dot{r}^2\,H' (r)],
\end{split}
\end{equation}
\begin{equation}\label{eq27}
\ddot{r}=\frac{1}{2\,c_0\,r}\,[c_2\,H (r)\,\dot{\psi}\,(2\,c_0\,r-r\,\dot{\psi})-c_0\,(\dot{r}^2+T\,\dot{\psi}^2-2\,c_1\,z\,\dot{\psi}\,\dot{z}-\dot{z}^2+2\,c_0\,\dot{\psi}\,\dot{T})].
\end{equation}

For the radial geodesic $\dot{z}=0=\dot{\phi}$. We have chosen $z=const$-planes defined by $z=z_0=0$, from eq. (\ref{eq26}) and (\ref{eq27}) we have
\begin{equation}\label{eq28}
\begin{split}
\ddot{T}&= \frac{1}{2\,c^{2}_0\,r}\,[-2\,c^{2}_{0}\,\dot{r}\,\dot{T}+c_0\,c_2\,H (r)\,\dot{r}^2+2\,c_0\,c_2\,r\,\dot{r}^2\,H' (r)],\\
\ddot{r}&= -\frac{\dot{r}^2}{2\,r}.
\end{split}
\end{equation}
The solution for $r$ gives
\begin{equation}\label{eq29}
\dot{r} (s)=\frac{c_4}{\sqrt{r}}\Rightarrow r (s)=[\frac{3}{2}\,(c_4\,s+c_5)]^{\frac{2}{3}},
\end{equation}
where $c_4, c_5$ are constants. From eq. (\ref{eq28}) we have
\begin{equation}\label{eq30}
\ddot{T}+\frac{\dot{r}}{r}\,\dot{T}=\frac{c_2\,c^{2}_4}{2\,c_0\,r}\,(2\,H' (r)+\frac{H (r)}{r}).
\end{equation}
From eqn. (\ref{eq30}) one can easily check for the chosen function $H(r)$ that the radial geodesics path $T$ is bounded for finite value of the affine parameter $s$ including $s=0$ ( since $r (s=0)=const \neq 0$ provided $c_5\neq 0$) and thus the study solution is radially geodesically complete.

To determine the strength of naked singularities, we consider the criterion developed in Refs. \cite{Tipler,Kro}. A sufficient condition for the singularity to be strong \cite{Tipler,Clarke2} is that
\begin{equation}\label{eq31}
\lim_{s\rightarrow 0} s^2\,R_{\mu\nu}\,\frac{dx^{\mu}}{ds}\,\frac{dx^{\nu}}{ds}\neq 0(>0),
\end{equation}
where $\frac{dx^{\mu}}{ds}$ is the tangent vector to the radial geodesics, and $R_{\mu\nu}$ is the Ricci tensor. While
the weaker condition, which we called the {\it limiting focusing condition} \cite{Kro} is defined by
\begin{equation}\label{eq32}
\lim_{s\rightarrow 0} s\,R_{\mu\nu}\,\frac{dx^{\mu}}{ds}\,\frac{dx^{\nu}}{ds}\neq 0.
\end{equation}

For the study solution (\ref{eq7}) we have
\begin{equation}\label{eq33}
\begin{split}
&\lim_{s\rightarrow 0} s^2\,[R_{rr}\,(\frac{dr}{ds})^2+R_{\psi\psi}\,(\frac{d\psi}{ds})^2]\\
&= \lim_{s\rightarrow 0} \frac{3\,s^2}{2\,r^2}\,\dot{r}^2,\quad \mbox{since}\quad \dot{\psi}=0\\
&= \frac{2\,c_4^{2}}{3}\,\lim_{s\rightarrow 0}[\frac{s^2}{(c_5+c_4\,s)^{2}}]\\
&= 0,\quad \mbox{if} \quad c_5\neq 0\\
&= const,\quad \mbox{if} \quad c_5=0.
\end{split}
\end{equation}
Similarly, the study solution does not satisfy the {\it limiting focusing condition}. Thus the naked singularities (NS) that is formed due to divergence of the scalar curvature does not satisfy neither the {\it strong curvature condition} nor the {\it limiting focusing condition} provided we assumed the constant $c_5>0$, otherwise only the strong curvature condition holds good for $c_5=0$. In that case ($c_5=0$) the study solution is geodesically incomplete.

\subsection{\bf The generalization of Misner space metric and Petrov classification}

We show below that the study solution (\ref{eq7}) is a four-dimensional generalization of Misner space metric in curved space-time. Before that we discuss the Misner space metric in 2D given by \cite{Mis}
\begin{equation}\label{eq34}
ds_{Mis}^2=-2\,dT\,d\psi-T\,d\psi^2,
\end{equation}
where $-\infty< T <\infty$ and $\psi$ coordinate is periodic. The Misner space metric in 2D is a locally flat space and regular everywhere. The curves defined by $T=T_0>0$, where $T_0$ is a constant being time-like and closed on account of periodicity of $\psi$, are formed closed time-like curves (CTC). The null curve $T=T_0=0$ serve as the Chronology horizon (since $g^{TT}=0$ at $T=T_0=0$) which divided the space-time a Chronal region without CTC to a non-chronal region with CTC. There is a Cauchy horizon at $T=const=T_0=0$ for any such space-like $T=const=T_0<0$ hypersurface (since $g^{TT}=T<0$ at $T=T_0<0$). Hence the space-time evolves from an initial space-like $T=const=T_0<0$ hypersurface in a causally well-behaved manner, up to a moment, {\it i. e.}, a null hypersurface $T=T_0=0$, and the formation of CTC takes place from causally well behaved initial conditions. The Misner space metric in 2D is a prime example of space-time where closed time-like curves develop at some particular moment. Levanony {\it et al} \cite{Leva} generalized this 2D flat Misner space in three and four-dimensional flat space. 

For constant $r,z$, from the metric (\ref{eq7}) we get (finally dropping bar)
\begin{equation}\label{eq35}
ds^2= r\,(-2\,c_0\,dT\,d\psi-T\,d\psi^2)=\Omega\,ds^{2}_{Misn},
\end{equation}
a conformal Misner space metric where, $\Omega=r$ is the conformal factor and we have done a transformation $T\rightarrow c^{-2}_0\,{\bar T}$, and $\psi \rightarrow c_{0}\,{\bar \psi}$.

To study the Petrov classification of the study solution (\ref{eq7}), one can construct a set of tetrad vectors $({\bf k,l,m,{\bar m}})$ \cite{Steph}. This set is given by
\begin{equation}\label{eq36}
\begin{split}
k_{\mu}&= \left (0, 0, 1, 0\right),\\
l_{\mu}&= r\,\left(c_0,-c_2\,H (r),\frac{T}{2},-c_1\,z \right ),\\
m_{\mu}&= \sqrt{\frac{r}{2}}\,\left(0,1,0,i \right ),\\
\bar{m}_{\mu}&= \sqrt{\frac{r}{2}}\,\left(0,1,0,-i \right ),
\end{split}
\end{equation}
where $i=\sqrt{-1}$. The set of tetrad vectors above is such that the metric tensor for the line element (\ref{eq7}) can be expressed as
\begin{equation}\label{eq37}
g_{\mu \nu}=-k_{\mu}\,l_{\nu}-l_{\mu}\,k_{\nu}+m_{\mu}\,\bar{m}_{\nu}+\bar{m}_{\mu}\,m_{\nu},
\end{equation}
where $-k^{\mu}\,l_{\mu}= m^{\mu}\,{\bar m}_{\mu}=1$, others are all vanishing. 

Using the set of null tetrad vectors (\ref{eq36}) we have calculated the five Weyl scalars. These are given by
\begin{equation}\label{eq39}
\Psi_0=\Psi_1=0=\Psi_2=\Psi_3,\quad \Psi_{4}=-\frac{1}{4\,c_0}\,[c_1-c_2\,H' (r)].
\end{equation}
Physically, the non-zero Weyl scalars $\Psi_4$ denotes a transverse wave components propagating along the principal null direction ${\bf k}$ of multiplicity 4. Therefore the considered null vector Eq. (\ref{eq36}) is aligned with the principal null directions (PND) along which the radiation propagates. In addition, the Weyl tensor $C_{\mu\nu\rho\sigma}$ satisfies the following Bel criteria, {\it i. e.,} 
\begin{equation}\label{eq40}
C_{\mu\nu\rho\sigma}\,k^{\sigma}=0.
\end{equation}
Thus the metric (\ref{eq7}) is of type N in the Petrov classification scheme. The complex scalar quantities $\Phi_{AB}={\bar \Phi}_{AB}$, $A,B=0,1,2$ associated with the Ricci tensor $R_{\mu\nu}$ are
\begin{eqnarray}
\Phi_{02}&=&\frac{1}{2}\,R_{\mu\nu}\,m^{\mu}\,m^{\nu}=\frac{1}{2}\,\rho,\nonumber\\
\Phi_{11}&=&\frac{1}{4}\,R_{\mu\nu}\,(k^{\mu}\,l^{\nu}+m^{\mu}\,{\bar m}^{\nu})=\frac{1}{4}\,\rho,\nonumber\\
\Phi_{22}&=&\frac{1}{2}\,R_{\mu\nu}\,l^{\mu}\,l^{\nu}=\frac{1}{4\,c_0}\,[c_1+c_2\,\{ \frac{H (r)}{r}+H' (r) \}],
\label{2019}
\end{eqnarray}
while others are all vanishing. The null vector $k_{\mu}$ satisfy the geodesics condition $k_{\mu\,;\,\nu}\,k^{\nu}=0$ with
\begin{equation}\label{eq41}
\boldsymbol{\Theta}=\frac{1}{2}\,k^{\mu}_{\,;\,\mu}=0,\quad \boldsymbol{\omega}^2=\frac{1}{2}\,{k_{[\mu\,;\,\nu]}}\,k^{\mu\,;\,\nu}=0,\quad \boldsymbol{|\sigma|}^2=\frac{1}{2}\,{k_{(\mu\,;\,\nu)}}\,k^{\mu\,;\,\nu}-\boldsymbol{\Theta}^2=0.
\end{equation}
But this null vector field ${\bf k}$ is not a covariantly constant vector field (CCNV) {\it i. e.,}
\begin{equation}\label{eq42}
k_{\mu\,;\,\nu}\neq 0.
\end{equation}
That means the study space-time (\ref{eq7}) admits a non-expanding, non-twisting, shear-free null vector field which is geodesic. This geodesic null congruence is not a covariantly constant null vector field and therefore, the study space-time exhibit geometrically different properties than the known {\it pp}-waves or plane-waves. 

\section{Conclusions}

In this paper, a family of type N exact non-vacuum solution to Einstein's field equations of zero cosmological constant satisfying the different energy conditions is studied. This family of a solution is regular everywhere except on the symmetry axis, where it possesses a naked curvature singularity. In sub-section {\it 2.1}, the stress-energy tensor of this solution is of the anisotropic fluid coupled with pure radiation field satisfies the different energy condition is presented. The physical parameters the fluid energy-density ($\rho$), the radial pressure ($p_{r}$) and the tangential pressures ($p_{t}$) diverge on the symmetry axis $r=0$. The radiation energy-density may be constant or function of spatial distance (r) for the chosen function $H (r)$ satisfied the null energy condition. After that, few type N and type O solution by Examples {\it 1}--{\it 4}, special sub-cases of the study solution is presented. In sub-section {\it 2.2}, we study the geodesic completeness of the type N solution. We have seen that this solution is radially geodesically complete in the $z=const$-plane provided $c_5>0$. Also, we study the strength of the naked singularity which is present due to the divergence of scalar curvature invariants  constructed from the Riemann tensor $R_{\mu\nu\rho\sigma}$ and/or divergence of the fluid energy-density ($\rho$). We have shown that the naked singularities satisfy neither the {\it strong curvature condition} given by Tipler \cite{Tipler} nor the {\it limiting focusing condition} by Krolak \cite{Kro} provided we have assumed the arbitrary parameter $c_5>0$, otherwise holds good only the strong curvature condition. In sub-section {\it 2.3}, we have shown that the study solution is a four-dimensional generalization of the Misner space metric in curved space-time admitting closed time-like curves which appear after a certain instant of time, a time-machine space-time. Furthermore, by constructing a set of null tetrad vectors we classify the study solution. We have seen that the non-zero Weyl scalars are only $\Psi_4\neq 0$ and rest are all vanishes. Also, the Weyl tensor satisfies the Bel criteria, namely, $C_{\mu\nu\rho\sigma}\,k^{\sigma}=0$, where $k^{\mu}$ is the null vector field aligned with the principal null directions (PNDs) along which the radiation propagates. Thus we have confirmed that the study solution is clearly of type N in the Petrov classification scheme. The study family of type N solution admits a non-expanding, non-twisting, and shear-free geodesic null congruence. But this shear-free geodesics null vector field is not a covariantly constant vector field and thus the rays of the gravitational waves are not parallel {\it i. e.,} the study type N solution exhibit geometrically different properties from the known {\it pp}-waves or plane-waves. 

The known special class of type N Kundt metrics with or without cosmological constant are either {\it pp}-waves or plane-waves and/or different from both of these ({\it e. g.}, Siklos solution and its sub-case). In this work, we have seen that the study family of type N non-vacuum space-time exhibit geometrically different properties from the known {\it pp}-waves, and belongs to a special class of type N Kundt metrics. Besides, this family of type N solution is a four-dimensional generalization of the Misner space metric in curved space-time, and behave as time-machine in the sense that the study space-time admits closed time-like curves appear after a certain instant of time. Therefore the study type N non-vacuum solution is different from the known class of Siklos solution. The study family of non-twisting type N non-vacuum solution with zero cosmological constant violates the causality condition and a special class of type N Kundt metrics. Furthermore, the time-dependent metric study here which is a four-dimensional generalization of the Misner space in curved space-time would represent a Cosmic Time Machine.
 
 \section*{Acknowledgement}
 Author sincerely acknowledge the anonymous kind referee(s) for their value comments and suggestions.
 
 \section*{Conflict of Interest:}
 The author declares that there is no conflict of interest regarding publication of this paper.
 
 \section*{Data Availability:}
 No data have been used to prepare this paper.

\end{document}